\documentclass[aps,prb,twocolumn,floats,showpacs,superscriptaddress]{revtex4-1}
\usepackage{graphicx,epsfig}
\usepackage{times,bbm}
\usepackage{graphics,dcolumn,bm,float}
\usepackage{amssymb,amsmath,rotate,color}
\usepackage[title,titletoc,toc]{appendix}
\usepackage{mathtools}
\usepackage{booktabs}

\usepackage[pagebackref=false,colorlinks,linkcolor=magenta,citecolor=cyan,urlcolor=magenta]{hyperref}

\begin{document}
\unitlength 1 cm
\newcommand{\be}{\begin{equation}}
\newcommand{\ee}{\end{equation}}
\newcommand{\bea}{\begin{eqnarray}}
\newcommand{\eea}{\end{eqnarray}}
\newcommand{\nn}{\nonumber}
\newcommand{\vk}{\vec k}
\newcommand{\vp}{\vec p}
\newcommand{\vq}{\vec q}
\newcommand{\vkp}{\vec {k'}}
\newcommand{\vpp}{\vec {p'}}
\newcommand{\vqp}{\vec {q'}}
\newcommand{\bk}{{\vec k}}
\newcommand{\bp}{{\bf p}}
\newcommand{\bq}{{\bf q}}
\newcommand{\br}{{\bf r}}
\newcommand{\bR}{{\bf R}}
\newcommand{\up}{\uparrow}
\newcommand{\down}{\downarrow}
\newcommand{\cdag}{c^{\dagger}}
\newcommand{\hlt}[1]{\textcolor{red}{#1}}

\title{Superconducting proximity in three dimensional Dirac materials: odd-frequency, pseudoscalar, 
pseudovector and tensor-valued superconducting orders}
\author{Zahra Faraei}
\thanks{Email: \href{mailto:zahra.faraei@gmail.com}{zahra.faraei@gmail.com}}
\address{Department of Physics, Sharif University of Technology, Tehran 11155-9161, Iran}
\author{S. A. Jafari}
\thanks{E-mail:   \href{mailto:jafari@physics.sharif.edu}{jafari@physics.sharif.edu}}
\address{Department of Physics, Sharif University of Technology, Tehran 11155-9161, Iran}
\address{Center of excellence for Complex Systems and Condensed Matter (CSCM), Sharif University of Technology, Tehran 1458889694, Iran}
\affiliation{Theoretische Physik, Universit\"at Duisburg-Essen, 47048 Duisburg, Germany}

\begin{abstract}
{We find that} a conventional s-wave superconductor in proximity to three dimensional Dirac material (3DDM), to all orders of perturbation in tunneling, induces a combination of s and p-wave pairing only. 
{We show that the} Lorentz invariance of the superconducting pairing prevents the formation of Cooper pairs with higher orbital angular momenta in the 3DDM.
This no-go theorem acquires stronger form when the probability of tunneling from the conventional superconductor to positive and negative energy states of 3DDM are equal.
In this case all the p-wave contribution except for the lowest order, identically vanish and hence we obtain an exact result for the induced p-wave superconductivity in 3DDM. 
Fierz decomposing the superconducting matrix we find that temporal component of the vector superconducting order and spatial components of the pseudo-vector order are odd-frequency pairing. {We find} that the latter is odd with respect to exchange of position and chirality of the electrons in the Cooper pair and is spin-triplet which is necessary for NMR detection of such an exotic pseudo-vector pairing. {Moreover, we show that} the tensorial order breaks into a polar vector and an axial vector and both of them are conventional pairing except for being spin-triplet. 
{According to our study,} for gapless 3DDM the tensorial superconducting order will be the only order which is odd with respect to the chemical potential $\mu$. Therefore {we predict that} a transverse p-n junction binds Majorana fermions.
{This effect can be used to control the neutral Majorana fermions with electric fields.}
\end{abstract}
\maketitle

\section {Introduction} 
Dirac equation combines relativistic and quantum aspects of the propagation of electron waves~\cite{Dirac1928,Dirac_Book}.
The charge conjugation symmetry of this equation has led to one of the landmark discoveries of the
20th century, namely the existence of anti-matter~\cite{ZeeBook}. 
In condensed matter, the Dirac equation emerges as an {\em effective} low-energy description of the band structure of a class of materials, called "Dirac materials" ~\cite{Wehling} ranging from graphene~\cite{NetoRMP} and helical conducting states on the surface of topological insulators~\cite{ShenBook} in two dimensions to tilted Dirac systems in organic systems~\cite{Tajima,Suzumura,Goerbig}, and more recent example of three-dimensional 
Dirac materials\cite{Armitage} such as (Mg, Al, Zn and Ca)BiSiO$_4$~\cite{Kane_BiSO4}. First principles DFT calculations showed that these materials are metastable and exhibit Dirac point degeneracies at T point of the Brillouin zone with no other band crossings at the Fermi level~\cite{Kane_BiSO4}. The necessary condition for a condensed matter system to allow for an effective description in terms of three dimensional Dirac equation is rather general. One only requires a small or vanishing gap, strong spin-orbit interaction and parity (P) plus time reversal (T) invariance~\cite{FuseyaReview,Kane_BiO2}. This set of conditions is basically the condensed matter statement of the CPT theorem, which means that the field of an electron must be invariant under the combined action of charge conjugation, parity and time reversal. In addition to the discrete symmetries of C, P and T, Dirac equation is covariant under the Lorentz transformation. When the 3+1 (space+time) dimensional version of Dirac equation comes into its mundane low-energy form in condensed matter systems, the velocity of light will be replaced by a velocity scale which is about 2-3 orders of magnitude smaller than the velocity of light. Breaking either P or T symmetry which maybe possible in some crystals, gives rise to the Weyl semimetals~\cite{WeylXu,WeylDai,Fang_Dai}. 

Once one has a 3DDM at hand, the nice thing about such a condensed matter realization is that one can bring it close to other interesting ground states of condensed matter, such as the superconducting state, and study the interplay between the Dirac nature of the wave equation in 3DDM and the proximity induced superconductivity. The standard superconducting proximity tells us that when a conventional BCS superconductor is brought next to a normal metal, the only form of superconductivity that can be induced in the metallic state is a conventional, spin-singlet, s-wave superconductivity. 
However, in this work we show that when a conventional s-wave superconductor is brought next to a 3DDM, much more interesting possibilities can arise. First of all, the strong spin-orbit coupling encoded in the very nature of the Dirac equation in 3DDM, allows us to have spin-triplet superconductivity as well~\cite{Narlikar,Bennemann}. This can be intuitively thought of as a Cooper pair that tunnels into the 3DDM and one of the electrons may (or may not) flip its spin~\cite{Salehi} due to the strong spin-orbit coupling in the 3DDM. 
This is in some sense the three dimensional generalization of the
Fu-Kane proposal where the induced superconductivity into the two-dimensional Dirac cone at the surface of a topological insulator permits triplet pairing~\cite{Fu2008PRL}. However in the present case we get much more than the two-dimensional case:
The two (space) dimensional Dirac equation on the surface of topological insulators is expressed in terms of
$2\times 2$ matrices which leave no room for a $\gamma^5$ matrix, and therefore the superconducting order parameters do not have any chance of becoming a {\em pseudo}-scalar, or {\em pseudo}-vector in the sense of transforming like $\gamma^5$ or $\gamma^5\vec\gamma$. However in the case of three (space) dimensional Dirac equation which is expressed in terms of $4\times 4$ matrices, there always exists a $\gamma^5$ matrix. Therefore the spin-singlet Cooper pairs induced into the 3DDM, can be either scalar or pseudo-scalar. Similarly the spin-triplet Cooper pairs induced into the 3DDM can be both vector and pseudo-vector. {The {\em pseudo} character for an order parameter implies that it changes its sign under mirror reflection and is therefore a Lorentz version of the odd-parity superconductivity. Here we describe how these "left-right"-breaking characters arise when one considers the four (space+time) dimensional Dirac equation in the bulk of a 3DDM.
It turns out that the pseudo-scalar pairing can lead to generation of Majorana fermions without requiring triplet pairing~\cite{Salehi2016}.}
In addition to pseudo-scalar and pseudo-vector channels, there remains yet another exciting form of superconductivity which behaves as a tensor under the Lorentz transformation. This is a unique chance that appears only in 3DDM which emerges in low-energy effective theory where the point group symmetry is enlarged to a much larger group of Lorentz transformations. 
Therefore, the 3DDM can be thought of as a unique platform that allows for unconventional superconducting pairings to be induced by simply placing it next to an abundant conventional BCS superconductor.

In this paper we employ a tunneling formulation and Green's function method to calculate the induced superconductivity in a 3DDM. 
We calculate the leading order induced $4\times 4$ superconducting pairing from which we extract the scalar, vector, pseudo-scalar, pseudo-vector and tensor superconducting orders in each of the above channels. As for the spatial part of the Cooper pair wave-function we find that {\em only} s-wave and p-wave superconducting correlations can be induced in the 3DDM and this is true to {\em all orders of tunneling}.
The bulk states in a 3DDM can consist of an odd or even numbers of massless Dirac cones~\cite{BohmJungYang}.
In this work we focus on a 3DDM with a single Dirac cone and find that there are circumstances under which the perturbative treatment becomes exact. 

The paper is organized as follows: In section~\ref{secII} we lay down the formulation by reviewing the fundamental charge conjugation symmetry of the Dirac equation from which we construct the appropriate Nambu spinor.
In section~\ref{secIII} starting with the most general possible form of the tunneling matrix, we calculate the Green's function in the Nambu space from which we extract the superconducting matrix. In section~\ref{secIV} we classify the superconducting order in a 3DDM with single Dirac cones in terms of their transformation properties under the Lorentz group.

\section{The Dirac Bogoliubov- de Gennes equation for 3DDM}
\label{secII}
The Dirac equation in 3DDM emerges as an effective theory under a rather general condition which basically requires a small gap and a large spin-orbit interaction~\cite{FuseyaReview}.
Naive discretization of 3+1 dimensional Dirac equation for crystals implies that the Dirac cones have to come in pairs~\cite{Morandi}. 
However examination of the crystal symmetries revealed that depending on the representation of the parity operator, either pairs of Dirac nodes exist that are pinned to opposite momenta, or there are odd number of Dirac cones one of which then must be at the center of the lattice Brillouin zone~\cite{BohmJungYang}. 
Let us start by the isotropic single-Dirac cone -- the so called Wolff Hamiltonian which was historically derived by Wolff~\cite{Wolff} for Bismuth -- as a prototype of the 3DDM.
The isotropic Wolff Hamiltonian for 3DDM is
\be
H_{0D}(\bk)=
v\begin{bmatrix}
   mv                       &   i \hbar {\bk}.\vec\sigma  \\
- i \hbar {\bk} . \vec\sigma  &           -mv
\end{bmatrix},
\ee
where $v$ is the Fermi velocity that replaces the velocity of light in 3DDM, $m$ sets the gap energy scale as $2mv^2$, the vector $\hbar\bk$ is the momentum measured from the Dirac point and ${\vec\sigma}$ denotes three Pauli matrices. 
From this point we set $\hbar$ and $v$ equal to $1$ and will restore the constants whenever required. 
In order to identify this Hamiltonian by the standard Dirac Hamiltonian,
\be
 H_{0D}(\bk)= {\bk} . \vec\alpha +  m \beta,
\ee
we make the following choice for the $\gamma^\mu$ matrices~\cite{gammas}
\be
   \gamma^0= \tau_3 \otimes \mathbbm{1},~~~
   \vec\gamma= \tau_1 \otimes i \vec\sigma,
   \label{gammaWolff.eqn}
\ee
in terms of which we construct $\beta=\gamma^0$ and $\vec\alpha=\gamma^0\vec\gamma$.
The Clifford algebra for $\gamma^\mu$ matrices implies $\vec\alpha\beta=-\beta\vec\alpha$ and $-\vec\gamma^2=\vec\alpha^2=\beta^2=1$. Note that Pauli matrices $\vec\tau$ act on the space of conduction and valence bands,
while $\vec\sigma$ act on the spin space. 

In this way, the Dirac equation for a charge $-e$ electron is given by,
\bea
[i  \gamma^0 \gamma^j ( \partial_j - i e A_j ) +  m \gamma^0] \psi_e = \varepsilon_e \psi_e ,
\eea
where $\psi_e$ is the wave function of an electron with momentum $\bk$ and energy $\varepsilon_e=\sqrt{k^2 + m^2}$ close to the Dirac point.  
To construct the appropriate Nambu spinor, we need to find out the wave equation for holes.
When a Dirac material is heavily doped away from the Dirac node, such that the
inter-band processes are negligible, the concept of a hole is quite close to its
standard one-band version according to which the wave function of a hole is basically
complex conjugate of the corresponding electron wave function~\cite{note1}.
Within this framework the superconductivity 
can be built in to the form of Bogoliubov-de Gennes construction with Nambu spinors
given by $(\psi_{\bk},\psi^\dagger_{-\bk})$~\cite{Beenakker}. However when the Dirac
material is at its neutrality point, the Dirac equation has a charge conjugation symmetry,
hence for any (four component) electron wave-function $\psi_e$ satisfying
the Dirac equation at energy $\varepsilon_e$, there exists another four component 
hole (positron) wave function $\psi_h=MK\psi_e$ at energy $\varepsilon_h=-\varepsilon_e$ 
that satisfies the same Dirac equation. Here $K$ stands for complex conjugation and $M$ is a
$4\times 4$ matrix whose explicit form, depends on the representation being used.
Therefore if the superconducting Hamiltonian in a 3DDM is to respect the Lorentz 
symmetry, the hole part of the corresponding Nambu spinor must be 
given by $MK\psi$. 

In the work of Fu and Kane~\cite{Fu2008PRL} it was found that a conventional s-wave superconductor can induce p-wave pairing into the two-dimensional Dirac cone on the surface of a topological insulator, provided the chemical potential is larger than the superconducting pairing scale. In this work we do not restrict ourselves to large chemical potentials, and among the other parameter regimes, we are particularly interested in the $\mu=0$ limit. With this motivation, first we need to review how the charge conjugation operator (particle-hole transformation) as an authentic symmetry of the Dirac equation can be constructed.
To obtain an equation for a hole with opposite charge, one needs to complex conjugate the Dirac equation,
\bea
[-i  \gamma^0 \gamma^{j*} ( \partial_j + i e A_j ) +  m \gamma^0] \psi_e^* =  \varepsilon_e  \psi_e^*.
\eea
This is the Hamiltonian for a particle of mass $m$ and charge $+e$, i.e. a hole.  
But since $- \gamma^{j*}$'s  also satisfy the Clifford algebra they can be obtained from the original
choice of $\gamma^\mu$'s by a similarity transformation $- \gamma^{j*}= M^{-1} \gamma^j M$. The
matrix $M$ in the present representation, turns out to be 
$M=\gamma^0 \gamma^2$. Plugging in the wave equation for holes (positrons) and using Clifford
algebra gives, 
\bea
 [i  \gamma^0 \gamma^{j} ( \partial_j + i e A_j ) +  m \gamma^0] \ M \psi_e^* = - \varepsilon_e \ M  \psi_e^*.
\eea
This clearly identifies an opposite charge wave function $\psi_h=MK\psi_e$ where $K$ is the complex conjugation, that satisfies the Dirac equation at energy $\varepsilon_h=-\varepsilon_e$. 
So the covariant form of the {\em Dirac}-Bogoliubov-de Gennes (DBdG) equation will be,
\bea
\begin{bmatrix}
H_{0D}(\bk)      &                                 { \Delta}\\
\\
                         {\Delta}^\dagger &                                         -H_{0D}(\bk) 
\end{bmatrix}
\begin{bmatrix}
\psi_e (\bk)                             \\
 M \psi_e^*(-\bk) 
\end{bmatrix}
=\varepsilon 
\begin{bmatrix}
\psi_e (\bk)                              \\
 M \psi_e^* (-\bk) 
\end{bmatrix},\nn
\eea
where in the Dirac equation for hole a $\bk\to-\bk$ is performed as we are interested in Cooper pairs with zero center of mass momentum. The superconducting order parameter
$\Delta$ is now a $4\times 4$ matrix in the space of Dirac indices $\mu=0\ldots 3$.
Note that the transformation $\psi\to MK\psi$ can alternatively be absorbed into the BdG matrix upon which the lower Block of the Hamiltonian will look like $-M^{-1}H^*_{0D}(\bk)M$. This emphasizes the contrast of the present Dirac BdG equation with non-Lorentz covariant Dirac BdG equation where the matrix $M$ is basically set to unit for spinless particles, and set to $i\sigma_y$ for spin-1/2 particles. 
Obviously missing matrix $M$ (e.g. by setting it equal to unit matrix) is one way of breaking the
Lorentz invariance of the ensuing superconducting pairing state. There are many more ways to do so by setting $M$ equal to any other matrix. 
We assume that the superconducting pairing does not break the Lorentz invariance.
Doping away from the Dirac node is straightforward and only affects the diagonal part of the BdG Hamiltonian,
\bea
{\mathcal H}_{\rm D}=
\begin{bmatrix}
m {\bf \gamma}^0+k_{\mu} \gamma^0 \gamma^{\mu}-\mu & { \Delta}                                                            \\
{\Delta}^\dagger                                               & \mu -m {\bf \gamma}^0- k_{\mu} \gamma^0 \gamma^{\mu }
\end{bmatrix}.
\label{H_DBdG}
\eea
Now let us explicitly construct the Nambu spinor, $\psi^\dag=[{\psi_e}^\dag(\bk)  ~,~ {\psi_h}^\dag(\bk)]$ where for a 3DDM,
\bea
\psi_e^{T} (\bk)=\big[
c_{\bk,+,\up}       ~
c_{\bk,+,\down}   ~
c_{\bk,-,\up}       ~
c_{\bk,-,\down}  \big],
\eea
and
\bea
\psi_h^{T}(\bk)&=& (M K\psi_e (-\bk))^T\\
\nn
&=&K\big[
c_{-\bk,-,\down}       ~
- c_{-\bk,-,\up}   ~
- c_{-\bk,+,\down}       ~
c_{-\bk,+,\up}   
\big],
\eea
in which, the subscripts  $\pm$ are orbital indices and refer to upper and lower bands with dispersin $\varepsilon=\pm\sqrt{k^2 + m^2}$.

\section{Proximity with an s-wave superconductor}
\label{secIII}
To develop our ideas let us bring a conventional s-wave superconductor, characterized with a scalar superconducting gap $\Delta_{sc}$, next to the 3DDM. We assume that $\Delta_{sc}$ is larger than the Dirac energy scale $m$.
We consider a planar interface perpendicular to the 
$z$-axis, located at $z=0$, with S (superconductor) region for $z<0$ and D (3DDM) region for $z>0$.
Combining the Nambu space of both the superconductor and the 3DDM, the Green's function in the absence of tunneling is given by
\bea
{\bf G}_0=
\begin{bmatrix}
G_{0S}    &       0      \\
    0                  &    G_{0D}
\end{bmatrix},
\eea 
where $G_{0S(D)}=[i\omega_n - \mathcal H_{S(D)}]^{-1}$ is a $4\times 4$ ($8\times8$) Green's function matrix in the Nambu space of the conventional superconductor (Dirac material). $\omega_n$ are Matsubara frequencies, $\mathcal H_S$ is the standard BCS Hamiltonian and $\mathcal H_D$ has been introduced in Eq~\eqref{H_DBdG}.

When the superconductor and 3DDM are brought together, the coupling between the two in the combined Nambu space can be described by a $4\times 8$ tunneling matrix ${\bf t}$.
This matrix has two blocks, one for the electron tunneling ( $\tau_e$) and the other for the holes ($\tau_h$).
The elements of $\tau_e$ ($\tau_h$) connect an electron (hole) annihilation operator from the Dirac material to an electron (hole) creation operator in the superconductor. The tunneling matrix given by:
\bea
\label{tildet.eqn}
{\bf t}=
\sum_{\langle \bk,\bk'\rangle} e^{-i(\bk'-\bk).{\vec r}} 
\begin{bmatrix}
   \tau_e                    &         0            \\
 0                   &               \tau_h              
\end{bmatrix}.
\eea
Here $\sum_{\langle\bk,\bk'\rangle}$ denotes the summation over $\bk$ and $\bk'$ with the limitation $\bk_{||}=\bk'_{||}$, where $||$ means parallel to the interface. Indeed, we consider a system with an interface parallel to the $xy$-plane, so $p_x=\hbar k_x$ and $p_y= \hbar k_y$ are good quantum numbers and remain unchanged through the tunneling process.
$\tau_e = (t_+ ~~ t_-)\otimes \mathbbm{1}$, as mentioned, describes the electron transfer from the 3DDM side to the superconductor, and $\tau_h= (t_- ~~ t_+) \otimes\mathbbm{1} $ represents the hole tunneling matrix, provided that the spin direction remains unchanged. Here $t_+$ and $t_-$ are the spin-independent tunneling amplitudes to positive and negative energy states of the 3DDM. 

The Green's function of 3DDM gets dressed at each order of tunneling and acquires off-diagonal matrix elements (${\bf F}_n$) in the Nambu space which are anomalous Green's functions and correspond to induced superconducting correlations.

\subsection*{Appearance of p-wave superconductivity}
To second order in tunneling, the Cooper pair propagator is,
\bea
\nn
{\bf F}_2 ={\textit g} (i\omega_n + \mu+ m  \gamma^0 +{\bk}. \vec\alpha) \mathcal{T} ( i\omega_n - \mu- m \gamma^0 - {\bk}. \vec\alpha)\\
\label{F2}
\eea
where $\mathcal{T}=\tau_e^{\dagger} \tau_h$ and
\bea
  {{\textit g}=\frac{m^* \pi}{\sqrt{\omega_n^2 + \Delta_{sc}^2}} \bigg[\frac{ e^{-\kappa_+ z}}{\kappa_+} - \frac{ e^{-\kappa_- z}}{\kappa_- }\bigg](\omega_n^2 + m^2 + k^2)^{-2}},\nn
\eea
{with $\kappa_{\pm}=\sqrt{{k_{||}}^2 \pm 2 i m^*({\omega_n}^2 + {\Delta_{sc}}^2)^{1/2}}$} resulting from integration over the $k'_z$ 
in the superconductor side. The $m^*$ is the effective mass in the superconductor defining the dispersion of
underlying band structure by $({k'}_x^2+{k'}_y^2+{k'}_z^2)/(2m^*)$ and ${k_{||}}^2=k_x^2+k_y^2$. The above factor is even function
of $\vk$ and $\omega_n$ and will not affect our symmetry considerations regarding 
the even/odd behavior under  space ($\vk\to -\vk$) and time reversal. 
Eq~\eqref{F2} can be written as,
\bea
{\bf F}_2 / {\textit g}
= A + B ({\bk}. \vec\alpha) + ( {\bk}. \vec\alpha)  \tau^{+\dagger} \tau^- ( {\bk}. \vec\alpha),
\eea
where $A= (i\omega_n+\mu+ m  \gamma^0 ) \mathcal{T} (i\omega_n- \mu- m  \gamma^0 )$
and $B$ are $4\times 4$ 
matrices independent of $\bk$. Therefore the first term of the above equation corresponds to s-wave 
superconductivity while the $ B ({\bk}. \vec\alpha)$ generates angular dependence proportional to 
$k_z$ and $k_x \pm i k_y$ which are $\ell=1$ spherical harmonics and hence corresponds to p-wave pairing.
Now let us focus on the third term which appears to be second order in $\bk$ and hence in general
is expected to mix d-wave harmonics. However the very structure of $ \tau_e^{\dagger} \tau_h$
decides about the fate of this term and higher order terms, which in our case this matrix is given by,
\bea
{\mathcal T} = 
\begin{pmatrix}
 t_+ t_-    &      0        &  t_+^2    &      0          \\
     0         &   t_+ t_-  &        0          &  t_+^{2}    \\
 t_-^{2}   &    0          &   t_+ t_-      &      0            \\
     0         &  t_-^{2}  &     0            &     t_+ t_-
\end{pmatrix}.
\label{Tonecone.eqn}
\eea 
This matrix describes the form of the tunneling matrix that has been appropriately folded
into the off-diagonal part of the Nambu space of the 3DDM.
It is now very useful to expand the above matrix in terms of a basis that is composed of 
one $\mathbbm{1}$, four $\gamma^\mu$, one $\gamma^5$, four $\gamma^5\gamma^\mu$, and 
six $\sigma^{\mu\nu}=i\gamma^\mu\gamma^\nu$ with ($\mu\ne\nu = 0,1,2,3$) and $\mu\ne\nu$. A general $4\times 4$ matrix 
$L$ can be expanded in this basis as,
\be
L=L_s\mathbbm{1}+L_5\gamma^5+L_\mu\gamma^\mu+L_{5\mu}\gamma^5\gamma^\mu+L_{\mu\nu}\sigma^{\mu\nu},
\label{gammaexpand.eqn}
\ee
where the indices have definite meaning with respect to Lorentz transformations: $L_s$ is
scalar, $L_\mu$ is vector, $L_5$ is pseudo-scalar, meaning that it is scalar except for 
transformations whose determinant is $-1$, e.g. mirror reflection. Similarly $L_{5\mu}$ is a pseudo-vector, 
and finally $L_{\mu\nu}$ is a rank two asymmetric tensor~\cite{ZeeBook}. 
This decomposition for ${\mathcal T}$ gives,
$$
{\mathcal T}= t_+t_- \mathbbm{1} - (i/2)\times[(t_+^2 - t_-^2)\gamma^5 - (t_+^2 + t_-^2)\gamma^5\gamma^0].
$$
With the commutation rules of the $\gamma$ matrices it can be seen that the above matrix 
can be manipulated as follows:
${\mathcal T}(\bk.\vec\alpha)=(\bk.\vec\alpha) \tilde{\mathcal T}$ where $\tilde {\mathcal T}$ is obtained from ${\mathcal T}$ by flipping the sign of the 
coefficient of  $\gamma^5 \gamma^0$. Hence the third term becomes,
\bea
  (\bk.\vec\alpha){\mathcal T}(\bk.\vec\alpha)= k^2\tilde{\mathcal T}
\eea
which means that the third term in the second order contribution
is also s-wave. 

At this point let us emphasize the importance of matrix $M$ required in the charge
conjugation: In the absence of matrix $M$, instead of $k^i k^j\alpha_i\alpha_j\tilde{\mathcal T}$ 
which was produced upon commuting $\bk.\vec\alpha$ to the left of tunneling matrix, we
would have $k^i k^j\alpha_i\alpha^*_j\tilde{\mathcal T}$. But unlike $\alpha_i\alpha_j$ tensor
which has a fully isotropic symmetric part,
the $\alpha_i\alpha^*_j$ does not have such an isotropic symmetric part and hence in addition to s-wave
component, d-wave component (which is of course compatible with singlet pairing)
would already appear at the lowest order of the tunneling. 

As can be inferred, the appearance of spin-triplet pairing is a result
of the spin-orbit interaction encoded in the form of $\bk.\vec\alpha$ in the Dirac Hamiltonian.
If we had started with a normal metal whose Hamiltonian is $k^2/2m_e$ (times the
unit matrix $\sigma_0$ in the spin space), there would be no Pauli spin matrices of spins involved, 
and proximity to s-wave superconductor would only induce s-wave pairing. However in the case of Dirac
Hamiltonian, the spin-orbit interaction inherent in $\bk.\vec\alpha$ structure generates higher spherical Harmonics, 
but then the $\{\alpha_i , \alpha_j \}= 2 \delta_{ij}$ structure (associated
with matrix $M$ and hence Lorentz invariance pairing) is responsible for cutting off the angular momenta hierarchy beyond the $\ell=1$. 

Let us see how this structure is preserved to {\em all orders of perturbation in tunneling}.
If we continue to calculate the higher orders, we find that the pairing potential is in general
made up of some powers of four kinds of terms: $\zeta^+ (\tau_{e/h}^\dagger \tau_{h/e}) \zeta^-$ 
and $\zeta^+ (\tau_{e/h}^{\dagger} \tau_{e/h}) \zeta^+$,  where 
$\zeta^\pm=i\omega_n\pm \mu \pm m  \gamma^0 \pm {\bk}. \vec\alpha$.
The two other terms are obtained by the permutation $\zeta^+\leftrightarrow\zeta^-$.
The lucky situation that happens here is that the set of matrices 
$\tau_{e/h}^\dagger \tau_{e/h}$  and $\tau_{e/h}^\dagger \tau_{h/e}$  
form a subgroup of $4\times4$ matrices of the form
$T=a\mathbbm{1}+a_0\gamma^0+a_5\gamma^5+a_{50}\gamma^{50}$. This is a subgroup as it is
closed under matrix multiplication. 
The interesting property of this group of matrices
is that when such a matrix passes through each $\bk.\vec\alpha$ (i.e. from the left of 
$\bk.\vec\alpha$ to its right) the expansion coefficients $a_m$ characterizing $T$ undergo
the transformation $(a_0,a_{50})\to -(a_0,a_{50})$ in the above expansion. 
Repeating this process to push all the tunneling matrices to the right, collects all the
$\bk$ dependence to the left, and we are eventually left with the elementary calculation of 
$(\bk.\vec\alpha)^m$ which is $k^m$ for even $m$ and $k^m(\hat k.\vec\alpha)$ for odd $m$ 
and hence at the end, we are left with a term proportional to 
$[ A_n + B_n ({\bk}. \vec\alpha) ] {\mathcal O} ({t_{\pm}}^{ 2 n})$ where $A_n$ and $B_n$ 
are some $\bk$-independent matrices. Therefore Lorenz invariance of the pairing which has been built into the 
matrix $M$ combined with the group property of the tunneling processes considered here, 
allows for the induction of {\em precisely} s- and p-wave superconductivity only and
prohibit the formation of higher angular momentum Cooper pairs.
Note that for weak links where tunneling amplitude is small, higher order tunneling processes 
are expected to become smaller in magnitude and hence contribute to a convergent series in 
the s- and p-wave induced superconductivity in a 3DDM. 

Indeed, so far we have convinced ourselves that when a conventional superconductor is placed
next to a 3DDM, the singlet Cooper pairs of the BCS superconductor can tunnel
into 3DDM either as spin-singlet or as spin-triplet Cooper pairs. 
But, in principle the spin singlet Cooper pair can correspond to any even angular momentum,
and the spin triplet Cooper pair would have corresponded to any odd angular momentum.
Our discussion establishes a no-go theorem according to which {\em only} s-wave
and p-wave angular momenta are possible. 

\subsubsection*{Exact results for a subset of 3DDM}
As will be discussed in materials sub-section~\ref{materials.sec}, the predicted candidate
materials for single cone 3DDM are expected to have nearly equal tunneling amplitudes from conduction and 
valence bands, namely $t_+\approx t_-$. With this motivation, let us study the special case where the tunneling amplitude 
corresponding to positive and negative energy states 
satisfy $t_+^2=t_-^2 =t^2$ which then further simplifies the tunneling matrix 
as ${\cal T}= t^2(\mathbbm{1}+i\gamma^{50})$. 

The set of matrices $x\mathbbm{1}+y\gamma^5\gamma^0$ forms a group that
by the very defining properties of $\gamma$ matrices, is isomorphic to the group of complex numbers
$z=x+iy$. This can be simply seen by assuming that if we are given two matrices 
${\cal T}_1=x_1\mathbbm{1}+y_1\gamma^5\gamma^0$ and ${\cal T}_2=x_2\mathbbm{1}+y_2\gamma^5\gamma^0$
parameterized by pairs of numbers $(x_1,y_1)$ and $(x_2,y_2)$, respectively, 
then their matrix product is given by 
${\cal T}_1{\cal T}_2=(x_1x_2-y_1y_2)\mathbbm{1}+(x_1y_2+x_2y_1)\gamma^5\gamma^0$,
which is precisely how two complex numbers $z_1=x_1+iy_1$ and $z_2=x_2+iy_2$ are
multiplied. Equipped with this observation, 
the property ${\cal T}(\bk.\vec\alpha)=(\bk.\vec\alpha)\tilde{\cal T}$ then can be
represented as $z(\bk.\vec\alpha)= (\bk.\vec\alpha)z^*$. Therefore even powers such as $[z(\bk.\vec\alpha)]^{2n}$
or equivalently $[z(\bk.\vec\alpha)z(\bk.\vec\alpha)]^n$ will become
$[(\bk.\vec\alpha)z^*z(\bk.\vec\alpha)]^n$ which is $(x^2+y^2)^n(k^2)^n$. For the odd powers
one has $\{z(\bk.\vec\alpha)\}^{2n+1}=(x^2+y^2)^nk^{2n}(\bk.\vec\alpha)z^*$.
For the above special case we have $x= t^2$ and $y= it^2$, the combination $x^2+y^2$ vanishes. 
In this case only the lowest order tunneling, i.e. $n=1$ survives and we have a 
stronger version of our no-go theorem: Still only s- and p-wave superconductivity
are induced in 3DDM, but in the symmetric tunneling case the whole contribution
of the p-wave pairing channel comes from the lowest order. Therefore {\em when the 
tunneling probability from conduction and valence bands of the 3DDM are the same,
essentially the lowest order result is exact.}
With this in mind, let us now make a detailed explanation about the induced superconductivity
in 3DDM.

\section{Classification of superconducting order in 3DDM}
\label{secIV}

So far, we showed that the Green's function of the 3DDM contains an 'anomalous' component, the pair amplitude $F_2$,  characteristic for superconducting systems~\cite{Belzig}.
In this section we are going to expand the induced superconducting pairing potential 
(from now on we use the symbol $\Delta$ instead of $F_2/\textit{g}$)
in "channels" -- in the sense of Eq.~\eqref{gammaexpand.eqn} -- with definite transformation 
properties under the Lorentz transformation. The spin-singlet induced pairing therefore breaks
into two pieces: it could either behave as a scalar, or a pseudo-scalar with respect to 
the Lorentz transformations. Similarly the spin-triplet induced pairing can have two 
components that transform either as a (four-) vector or as a pseudo (four-) vector.
In addition we could have a component which may behave as a tensor. 
The Fierz decomposition~\cite{Goswami} of the superconducting matrix will be
\bea
\Delta=\Delta^s +  \Delta_{\mu} \gamma^{\mu}  +  \Delta_{\mu \nu}  \sigma^{\mu \nu}+\Delta_{5\mu} \gamma^{5} \gamma^{\mu} + \Delta_5 \gamma^5
\label{deltaexpand.eqn}
\eea 
where the basis is defined in Eq.~\eqref{gammaexpand.eqn}.
 The resulting superconducting orders in various
”channels” are summarized in table~\ref{single_gap_dirac.tab}.

\begin{table} [t] 
\centering
\begin{tabular}{r  p{7.5 cm}}
  \hline
\hline
$ $ & $ $ \\ [-0.2cm]

$\Delta^s:$ & $ - t_+ t_- (\omega_n^2 + m^2 + \mu^2 + k^2) $ \\[0.2cm]

$\Delta_5:$ & $  \frac{i}{2} (t_+^2 - t_-^2) (\omega_n^2 - m^2 + \mu^2 + k^2) + \omega_n m (t_+^2 + t_-^2)   $ \\[0.2cm]

$\Delta_0:$ & $  2 t_+ t_- m \mu$ \\[0.2cm]

$\Delta_{50}:$ & $ \frac{i}{2} (t_+^2 + t_-^2) (\omega_n^2 - m^2 + \mu^2 - k^2) - \omega_n m (t_+^2 - t_-^2) $  \\[0.2cm]

$\Delta_{5j}:$ & $  i[m(t_+^2 - t_-^2) - i \omega_n (t_+^2 + t_-^2)] k_j$ \\[0.2cm]

$\Delta_{0j}:$     & $    2 \mu t_+ t_-  k_j $ \\[0.2cm]

 $\Delta_{ij}:$ & $  - i \mu (t_+^2 - t_-^2)  \epsilon_{ijl} k_l $  \\[0.1cm]
 \hline
\hline
\end{tabular}
\caption{Fierz decomposition of the gap matrix for 3DDM \label{single_gap_dirac.tab}. The indices $i,j,l$ correspond
to three spatial directions $1,2,3$ and $\epsilon_{ijl}$ is the totally antisymmetric tensor.}
\end{table}

On the other hand, it is crucial to note that the superconducting matrix (anomalous Green's function matrix)
is basically $\langle \psi_e\bar\psi_h\rangle$ where $\bar\psi_h=\psi^\dagger_h\gamma^0$. Using the fact that the hole is obtained by $\psi_h=MK\psi_e$ with $M=\gamma^0\gamma^2$, after some algebra we find that
\be
   \bar\psi_h=\psi_e^T M^\dagger\gamma^0=-\psi^T_e \gamma^2.
\ee
Using explicit representation of $\gamma$ matrices, and assuming
\be
\psi_e=
\begin{bmatrix}
c_{\bk,+,\up}\\ c_{\bk,+,\down} \\c_{\bk,-,\up}    \\ c_{\bk,-,\down}
\end{bmatrix},\nn
\ee
we explicitly obtain,
\be
   \bar\psi_h=\begin{bmatrix}
  c_{-\bk,-,\down}&- c_{-\bk,-,\up}& c_{-\bk,+,\down} & -c_{-\bk,+,\up}   
   \end{bmatrix},
\ee
which then using the definition 
$\Delta_{\alpha\sigma,\alpha'\sigma'}=\langle \psi_{e\alpha\sigma}\bar\psi_{h\alpha'\sigma'}\rangle$
with $\alpha,\alpha'=\pm$ and $\sigma,\sigma'=\up,\down$ gives the structure,
\bea
&\Delta  =
\begin{bmatrix}
\Delta_{+\up - \down}    & -\Delta_{+\up - \up}    & \Delta_{+\up + \down}    & -\Delta_{+\up + \up}\\
\Delta_{+ \down - \down} & -\Delta_{+ \down - \up} & \Delta_{+ \down + \down} & -\Delta_{+ \down + \up}\\
\Delta_{-\up - \down}    & -\Delta_{- \up - \up}   & \Delta_{-\up + \down}    & -\Delta_{-\up + \up}\\
\Delta_{- \down - \down} & -\Delta_{- \down - \up} & \Delta_{- \down + \down} & -\Delta_{- \down + \up}
\end{bmatrix}
\label{gapmatrix.eqn}.
\eea

Here we have two bands 
(labeled by orbital index $\alpha=\pm$) and two spin degrees of freedom ($\sigma=\up,\down$) which are all encoded in the 
gap matrix, Eq.~\eqref{gapmatrix.eqn}, giving a total of $16$  possible pairing amplitudes.
Some of them are inter-band and some are intra-band pairings. 
As discussed, the orbital angular momentum of the pairing function can only be 
s-wave or p-wave to all orders. The $\up\up$ or $\down\down$ total spin is proportional to
$k_x\pm ik_y$ while the $\up\down$ or $\down\up$ can either be proportional to $k_z$ (meaning
p-wave with $\ell=0$) or independent of angle (meaning an s-wave spin singlet pair). 
As can be seen, the values reported in table~\ref{gapmatrix1cone.tab} obey this form.

 \begin{table} [t]
\centering
\begin{tabular}{r|  p{7.1 cm} }
  \hline
\hline
$ $ & \\
$\Delta_{\alpha \sigma ,  \alpha \sigma}$  &  $ 2 i \alpha\sigma  \mu   t_+ t_- (k_x + i \sigma k_y)$ \\ [0.1cm]
\hline
$ $ &    \\
$\Delta_{\alpha \sigma ,  \bar{\alpha} \sigma}$ & $   i\alpha \sigma [ (m + \alpha \mu)(t_+^2 - t_-^2) - i \omega_n(t_+^2 + t_-^2)] (k_x + i \sigma k_y)$  \\[0.1cm]
\hline
$ $ &\\
$\Delta_{\alpha \sigma , \alpha \bar{\sigma} }$ &  $- \sigma[ t_{\bar{\alpha}}^2 (\omega_n^2 - m^2 - \mu^2) - t_\alpha^2 (k^2 - i \alpha \omega_n m)] - \alpha \sigma t_+ t_- k_z $\\[0.1cm]
\hline
$ $ &\\
 $\Delta_{\alpha \sigma , \bar{\alpha} \bar{\sigma}}$ &  $ -\alpha \sigma t_+ t_- [\omega_n^2 +  (m - \alpha \mu)^2+ k^2 ] $ \\
 $ $ & $  - i\alpha [  (m+ \alpha \mu) ( t_+^2 - t_-^2)  - i\omega_n (t_+^2 + t_-^2) ]k_z$   \\[0.1cm]
\hline
\hline
\end{tabular}
\caption{Values of the gap matrix elements in Eq~\eqref{gapmatrix.eqn} for a 3DDM.  
$\alpha$ and $\sigma$ refer to spin and band index, respectively. $\bar{\alpha}=-\alpha$ and $\bar{\sigma}=-\sigma$.
The band index $\alpha$ for $m=0$ coincides with the chirality label $\chi$ in table~\ref{symmetries.tab}.
\label{gapmatrix1cone.tab} }
\end{table}

Clearly table~\ref{single_gap_dirac.tab} and table~\ref{gapmatrix1cone.tab} are two different ways of representing the same superconducting correlations. Indeed the second column of table~\ref{symmetries.tab} presents the relation between these two ways of representing the superconducting correlations in a 3DDM which is obtained from Eq.~\eqref{deltaexpand.eqn} for the matrix in Eq.~\eqref{gapmatrix.eqn}. The third column of table~\ref{symmetries.tab} indicates the sign arising from the exchange of the spins of the two electrons in the Cooper pair amplitudes in the second column and is valid for arbitrary $m$. The fourth column corresponds to the sign change arising from the exchange of the band attribute, $\alpha$ of the electrons in the Cooper pair. For $m=0$ this coincides with the chirality attribute $\chi$. The projection to states with a definite chirality $R$ or $L$ is defined as $\psi_{R/L}=(1\pm\gamma^5)\psi/2$. Chiral states are eigen-states of either of these projections. At $m=0$, the states with definite band index $\alpha$ have definite chirality $\chi=\alpha$~\cite{ZeeBook,PeskinBook}.
The fifth column ($P$) is the parity eigenvalue of the Cooper pairing amplitude which arises from the transformation $\vk \to -\vk$ and is extracted from the second order tunneling results of table~\ref{single_gap_dirac.tab}.
Note that the parity of the $\Delta_j$ can not be extracted from table~\ref{single_gap_dirac.tab}
as it is zero at the present leading order. It can in principle appear in higher orders of perturbation theory. Its symmetry however can be deduced from the following argument: We would like to construct a vector function $\Delta_j$ of a vector $\vk$. Since the only vector in the problem is $\vk$, the gap function $\Delta_j$ can only be odd (parity) function of $\vk$. This argument indeed holds for those parts of the table where we have non-zero lowest order pairing. 
For example all triple entities such as $\Delta_{5j}$,  $\Delta_{0j}$ and $\Delta_{ij}$ being Cartesian components of pseudo-vector, polar-vector and axial vectors, respectively satisfy this property.

The sixth column can in principle be constructed from the requirement of total antisymmetry of the
Cooper pair amplitude under the exchange of {\em all} attributes of the electrons, i.e. their spin, chirality, position (parity), and time~\cite{blackShaffer}. As can be seen, the temporal component of the four-vector, namely $\Delta_0$ and spatial portion of the pseudo-four-vector, namely $\Delta_{5j}$, give rise to odd-frequency pairings~\cite{TanakaJPSJ}. The existence of odd frequency pairing is in agreement with earlier work on the possibility of odd frequency pairing in multi-band systems~\cite{blackShaffer}.
However for $m\ne 0$, as can be seen in table~\ref{single_gap_dirac.tab}, there appear terms proportional to $m$ that seems to violate the expectation from the 6th column of table~\ref{symmetries.tab}. 
This can be rooted back to the fact that the eigen-states of massive Dirac equation do not have definite chirality, and e.g. the positive energy eigen-states is dominated by right ($+$) chirality, except for a little bit mixing of left ($-$) chirality proportional to $m$~\cite{ZeeBook,PeskinBook}. 
 
{Therefore the even frequency contribution to $\Delta_0$ in table~\ref{single_gap_dirac.tab} actually arises from such a frequency mixing on top of a vanishing principal odd-frequency component. This is why in table~\ref{symmetries.tab} the principal odd-frequency which can be deduced from symmetry appears as "-".}

\begin{table} [t] 
\centering
\begin{tabular}{ | c |  c |  c|   c | c | c| }
\hline
$\Delta$ &{Cooper pairing}   & $S$ & $\chi$ & $P$ & $\omega$  \\
\hline
\cline{3-6}
$\Delta^s$ & $ \Delta_{+\up - \down} - \Delta_{+\down - \up} + \Delta_{-\up + \down} - \Delta_{-\down +\up} $ & $-$& $+$ & $+$& $+$ \\
\hline
$\Delta_5$ & $ \Delta_{+\up + \down} - \Delta_{+\down + \up} - \Delta_{-\up - \down} + \Delta_{-\down -\up} $ & $-$& $+$ & $+$& $+$ \\
\hline
$\Delta_0$ & $   \Delta_{+\up - \down} - \Delta_{+\down - \up} - \Delta_{-\up + \down} + \Delta_{-\down +\up}$ & $-$& $-$ & $+$& "$-$"\\
\hline
$\Delta_1$ &$ \Delta_{+\up + \up} - \Delta_{+\down + \down} + \Delta_{-\up - \up} - \Delta_{-\down -\down}$ & $+$& $+$ & "$-$"& $+$ \\
\hline
$\Delta_2$ &$ \Delta_{+\up + \up} + \Delta_{+\down + \down} + \Delta_{-\up - \up} + \Delta_{-\down -\down}$ & $+$& $+$ & "$-$"& $+$ \\
\hline
$\Delta_3$ &$  -\Delta_{+\up + \down} - \Delta_{+\down + \up} - \Delta_{-\up - \down} - \Delta_{-\down -\up}$ & $+$& $+$ & "$-$"& $+$ \\
\hline
$\Delta_{50}$ &$ \Delta_{+\up + \down} - \Delta_{+\down + \up} + \Delta_{-\up - \down} - \Delta_{-\down -\up}$ &$-$& $+$ & $+$& $+$ \\
\hline
$\Delta_{51}$ &$ \Delta_{+\up - \up} - \Delta_{+\down - \down} - \Delta_{-\up + \up} + \Delta_{-\down +\down}$ &$+$& $-$ & $-$& $-$ \\
\hline
$\Delta_{52}$ &$ -\Delta_{+\up - \up} - \Delta_{+\down - \down} + \Delta_{-\up + \up} + \Delta_{-\down +\down}$ & $+$& $-$ & $-$& $-$ \\
\hline
$\Delta_{53}$ &$ -\Delta_{+\up - \down} - \Delta_{+\down - \up} + \Delta_{-\up + \down} + \Delta_{-\down +\up}$ &$+$& $-$& $-$& $-$ \\
\hline
$\Delta_{01}$     & $ -\Delta_{+\up + \up} + \Delta_{+\down + \down} + \Delta_{-\up - \up} - \Delta_{-\down -\down} $& $+$& $+$ & $-$& $+$ \\
\hline
$\Delta_{02}$     & $ \Delta_{+\up + \up} + \Delta_{+\down + \down} - \Delta_{-\up - \up} - \Delta_{-\down -\down} $ & $+$& $+$ & $-$& $+$ \\
\hline
$\Delta_{03}$     & $ \Delta_{+\up + \down} + \Delta_{+\down + \up} - \Delta_{-\up - \down} - \Delta_{-\down -\up} $ & $+$& $+$ & $-$& $+$ \\
\hline
$\Delta_{12}$ & $ \Delta_{+\up - \down} + \Delta_{+\down - \up} + \Delta_{-\up + \down} + \Delta_{-\down +\up} $ & $+$& $+$ & $-$& $+$  \\
\hline
$\Delta_{23}$ & $ -\Delta_{+\up - \up} + \Delta_{+\down - \down} - \Delta_{-\up + \up} + \Delta_{-\down +\down} $ & $+$& $+$ & $-$& $+$  \\
\hline
$\Delta_{13}$ & $ -\Delta_{+\up - \up} - \Delta_{+\down - \down} - \Delta_{-\up + \up} - \Delta_{-\down +\down} $ & $+$& $+$ & $-$& $+$  \\
\hline
\end{tabular}
\caption{Pairing symmetries in 3DDM. The first column is the superconducting amplitude in various channels (scalar, pseudo-scalar, vector, pseudo-vector and tensor). The second column indicates the explicit expression for the Cooper pairs which is obtained by Fierz decomposition of 
Eq.~\eqref{gapmatrix.eqn} such that it satisfies Eq.~\eqref{deltaexpand.eqn}. Third column ($S$) indicates the sign arising from the exchange of the spins of electrons in a Cooper pair. Fourth column indicates the sign that arises from the exchange in the $+$ and $-$ (band) attributes of the electrons in the Cooper pair.
For $m=0$ this corresponds to exchange of chiralities ($\chi$).
Fifth column ($P$) indicates the sign that arises from $\vk \to -\vk$ in table~\ref{single_gap_dirac.tab}.
Although in the present second order perturbation result summarized in table~\ref{single_gap_dirac.tab} there are no $\Delta_j$ contributions, but since the only vector in the problem is $k_j$, the only acceptable functional dependence of $\Delta_j$ on $k_j$ can have odd parity. 
That is why we have used quotaton mark to indicate the putative parity (perhaps at higher orders of perturbation theory) of the
$\Delta_j$. Last column follows from total antisymmetry under exchange of all attributes which agrees with 
table~\ref{single_gap_dirac.tab} {\em only} for $m=0$. {For any deviation of $m$ from $0$, frequencies other than those indicated in this column can mix. At $m=0$, there would be no $\Delta_0$ in the leading order perturbation result of table~\ref{single_gap_dirac.tab}, but if anything appears in higher orders must be odd frequency. Any $m\ne 0$ mixes little bit of the opposite (i.e. even frequency) in agreement with table~\ref{single_gap_dirac.tab}. }
\label{symmetries.tab}}
\end{table}

This indicates that in the present classification of the superconducting order in 3DDM, the chirality $\chi$ 
is suitable rather than the orbital index $\alpha$. For the $m=0$ (gapless Dirac) situation, these two attributes
become identical and hence the even/odd frequency behavior expected from table~\ref{symmetries.tab}, agrees with
those obtained from the concrete tunneling calculation of table~\ref{single_gap_dirac.tab}.

\subsection*{Pseudo-scalar, pseudo-vector and tensor-valued superconductivity}
Having clarified the frequency behavior of the gap function in various channels in table~\ref{single_gap_dirac.tab},
let us now discuss in detail the contents of this table. 
With respect to scalar behavior under rotation, we have two possibilities:
(i) Lorentz-scalar superconducting order which is denoted by $\Delta^s$ and is the coefficient of the
matrix ${\mathbbm 1}$ in Eq.~\eqref{deltaexpand.eqn}.
(ii) The next possible order which belongs to spin-singlet Cooper pairing is 
the pseudo scalar superconductivity, $\Delta_5$ which is the coefficient of matrix $\gamma^5$
in expansion of the superconducting matrix, c.f. Eq.~\eqref{deltaexpand.eqn}. 
It can be confirmed from the second (and hence third) column of table~\ref{symmetries.tab} that these two
superconducting orders correspond to spin singlet Cooper pairs. 
There is a topological significance associated with the pseudo-scalar, $\Delta_5$ pairing:
Indeed we have recently shown that the order parameter $\Delta_5$, {\em despite being spin-singlet} can in competition 
with the Dirac gap $m$, itself give rise to a two dimensional sea of Majorana zero modes~\cite{Salehi2016}.
The present work shows that the pseudo-scalar superconductivity can be possibly obtained by 
proximity of a BCS superconductor to a 3DDM. {The recent observation of $4\pi$-periodic 
Andreev bound states~\cite{Brinkman} is a very strong evidence for existence of the pseudo-scalar
superconductivity.} 
Moreover, under the lucky circumstance of
almost equal tunneling amplitude to upper and lower bands, $t_+\approx t_-$, the present second order
result will be almost exact.
 
The next level of complexity in the superconducting order is the {\em four-vector}
superconducting order. The three-vector version of it, is familiar in the standard
triplet pairing context. However being a four-vector $(t,\vec r)$, the length $t^2-r^2$
of a four-vector can be positive corresponding to time-like separations, or negative
corresponding to space-like separations. With this brief reminder, let us now compare
the induced $\Delta_\mu$ orders parameters in 3DDM problem. As can be seen in 
table~\ref{single_gap_dirac.tab}, within the lowest order perturbation theory in 3DDM
only $\Delta_0$ is non-zero, and the spatial part $\Delta_j$ with $j=1,2,3$ is 
identically zero. This means that the pairing corresponding to (four) vector 
pairing in 3DDM is purely time-like. This is expected to show interesting properties
when an electric field and a magnetic field are applied together to such a superconductor.
One can imagine Lorentz transforming to a reference frame to eliminate the electric field $\vec E$~\cite{Goerbig2009}.
In such a system we will be dealing with the Meissner response of a superconductor where both 
$\Delta_0$ (spin-singlet, odd frequency) and $\Delta_i$ (spin-triplet, even-frequency) 
are non-zero.  

Within the Lorentz group, four-vectors can behave as pseudo-vectors, in the sense of being a coefficient of $\gamma^5\gamma^\mu$ in expansion~\eqref{deltaexpand.eqn}~\cite{machida2014}. These are denoted by $\Delta_{5\mu}$. In the case of 3DDM as can be seen in table~\ref{single_gap_dirac.tab}, both temporal and spatial components are non-zero. From table~\ref{symmetries.tab} for pure chiral pairing ($m=0$) we expect $\Delta_{5j}$ $(\Delta_{50})$ to be odd- (even-) frequency.
As pointed out, non-zero $m$ mixes a little bit of the opposite chirality in proportion to $m$ which then, as can be seen in table~\ref{single_gap_dirac.tab}, adds in an even (odd-) frequency contribution in proportion to $m$. The dominant odd-frequency pairing arises only for $\Delta_0$ and $\Delta_{5j}$.
Therefore we confirm the existence of odd-frequency pairing in two-band systems~\cite{blackShaffer} and in addition we identify this odd-frequency pairing as a pseudo-vector with respect to Lorentz transformation.
This pairing is spin-triplet, odd-chirality, and odd-parity. 
The proportionality of $\Delta_{5j}$ to $k_j$, nicely indicates its vector character with respect to space rotations. 

{ 
Let us see what is the essential property of pseudo-scalar $\Delta_5$ superconductivity as compared to the scalar $\Delta^s$ pairing: 
Imagine a transformation (a reflection) that changes the name of orbital indices $\pm$. This reflection maps the scalar $\Delta^s$ to itself, while $\Delta_5$ changes sign. 
Similarly as can be seen from table~\ref{symmetries.tab}, under the same operation, the spatial component of the vector order $\Delta_j$ (as e.g. in $^3$He superconductor) does not change sign, while the spatial components of the pseudo-vector, $\Delta_{5j}$ change sign. 
The physical content of such Z$_2$ form of a left-right symmetry breaking is no less than e.g. SO(3) symmetry breaking that spontaneously pick up a direction in space for a magnet, or U(1) symmetry breaking that picks a definite phase for a superconductor.
}

Finally at the highest level of complexity we have tensor superconducting order, $\Delta_{\mu\nu}$ for $\mu\ne\nu$. As can be seen in table~\ref{single_gap_dirac.tab},  the six tensorial components break into a polar vector $\Delta_{0j}\sim k_j$ and an axial vectors 
$\Delta_{ij}\sim \epsilon_{ijl}k^l$ where $i,j,l$ are the spatial indices $1,2,3$. 
All six components being grouped into vector will be spin-triplet (even), and since their $\bk$-dependence is odd. Since in table~\ref{symmetries.tab} they correspond to even-chirality pairing, they will correspond to normal even-frequency pairing. 
The interesting aspect of the tensorial part is that it vanishes as the chemical potential $\mu$ approaches the Dirac node. Particularly when $m=0$, the tensorial part will be the only superconducting pairing that scales with $\mu$ and {\em changes sign as $\mu$ does}.
This can be used to experimentally single out the contribution of tensor superconducting order.

This scenario becomes particularly interesting when a p-n junction is built in the transverse plane.
Across the junction (in the $xy$ plane) the tensorial superconducting order changes sign, and therefore the p-n junction is expected to bind Majorana fermions. Such Majorana fermions bound to lateral p-n junction is exclusively from tensorial superconducting order. 
{Let us see how does this come about: 
To get Majorana fermions in condensed matter systems, one simply needs two competing mechanisms to close and reopen a superconducting gap~\cite{Salehi2016, Beennaker}. Our leading order tunneling results show that when $m=0$, the tensorial part will be the only superconducting pairing. But on the other hand it scales with $\mu$ and therefore can change sign if $\mu$ changes sign. 
This allows us to conclude that if one constructs a lateral p-n junction with three dimensional Dirac materials, both p and n sides give rise to superconducting gaps of opposite signs and therefore there should be an interface region where the gap closes and hence the p-n junction is expected to bind Majorana fermions. The gap closes at the p-n interface between opposite gap signs and Majorana fermions will be confined to the interface. This opens up the possibility of electric-field control of Majorana fermions. Given that Majorana fermions are charge-neutral, the possibility to manipulate them by electric field is worth further investigations.
}

Furthermore the axial portion of the tensorial superconducting order is expected to display interesting Meissner effect as both $\Delta_{ij}$ and the electromagnetic field $\vec B$ are axial, and their coupling requires a pseudo-scalar coupling~\cite{Moore}. 
\subsubsection*{Materials}
\label{materials.sec}
The distorted spinel structure such as ZnBiSiO$_4$ has a single Dirac cone at T point~\cite{Kane_BiSO4}. The states near the Fermi surface are dominated by $p$-like states of the Bismuth atoms, so that, for the tunneling amplitudes we expect $t_-\approx t_+$. Here $m$ is zero and in an undoped 
($\mu=0 $) sample at zero temperature, as can be seen from table~\ref{single_gap_dirac.tab},  only $\Delta^s$ and $\Delta_{50}$ (both with spin-singlet, s-wave, even-chirality, even-frequency) along with
$\Delta_{5j}$ (spin-triplet, p-wave, odd-chirality, odd-frequency) will be non-vanishing. 
This simply means that an applied magnetic field can suppress $\Delta^s$ and $\Delta_{50}$ in favor of the exotic $\Delta_{5j}$ pairing. For NMR experiments, this implies that since in this particular case, spin-triplet pairing is locked to odd-chirality, odd-frequency and pseudo-vector character of $\Delta_{5j}$, the NMR signature of triplet pairing would be tantamount to: (1) the pairing in such materials is odd-frequency and (2) the pairing is odd with respect to exchange of chirality attributes, and (3) the pairing order parameter behaves as a pseudo-vector with respect to the Lorentz transformation. 
Moreover if one can tune $\mu$ away from zero, a new term $\Delta_{ij}$ emerges which is proportional to $\mu$. This should be contrasted to the existing $\Delta^s$ and $\Delta_{5j}$  terms which are proportional to $\mu^2$. 

\section{Summary}
In this paper, we have studied the induction of superconductivity from a conventional BCS superconductor to a 3DDM. 
First of all we make sure that the Nambu spinor is constructed from an electron and a "hole" that are precise charge conjugation of the Dirac operator. Moreover for the superconducting 3DDM we ensure the Lorentz covariance of the formulation, which essentially means being careful to use 
$\bar\psi=\psi^\dagger\gamma^0$ instead of simply $\psi^\dagger$ routinely used for non-Dirac condensed matter. 
This gives the peculiar arrangements of the pairing amplitudes as in Eq.~\eqref{gapmatrix.eqn}. 

For the most general form of tunneling matrix elements consistent with symmetries, and allowing for non-equal tunneling into valence and conduction states, we find that tunneling can be encoded into a set of matrices which form a subgroup of the Dirac matrices. This "tunneling group" property combined with the peculiar structure of the Dirac $\gamma$ matrices satisfying the Clifford algebra, to all orders in perturbation theory, allows only for the $\ell=0$ (s-wave) and $\ell=1$ (p-wave) orbital angular momenta, and higher angular momentum combination of the components of the vector $\bk$ can never be generated by higher orders of tunneling. 

Focusing on the explicit calculation of the induced superconductivity in the second order perturbation theory, the resulting expression for the superconducting matrix when decomposed as in Eq.~\eqref{deltaexpand.eqn} into various channels with definite transformation properties under the Lorentz transformation would give rise to a zoo of scalar, pseudo-scalar, four-vector, pseudo-four-vector, and tensorial superconducting order parameters. As for the symmetry of the pairing amplitude summarized in table~\ref{symmetries.tab}, we find that the appropriate attribute to classify the symmetry is the chirality rather than the band or orbital index. These two attributes coincide for $m=0$. The effect of non-zero $m$ is to mix a contribution from the state with opposite chirality. 
This is the root of mixed frequency behavior in table~\ref{single_gap_dirac.tab}.
When $m=0$, we get pure even- or pure odd-frequency (only in $\Delta_{5j}$) pairing.
The tensorial superconducting order splits into a polar and axial vector portions, every one of which scales with first power of the doping level $\mu$ measured from the Dirac node. 
Odd dependence on $\mu$ implies that a p-n junction in the transverse plane can bind Majorana fermions which are exclusively from tensorial superconducting pairing. 
This effect can be used as a platform to control Majorana fermions by electric fields. 

For the special case of $t_+=t_-$ -- which is a very good approximation for realistic 3D Dirac materials -- we showed that all higher order tunneling corrections identically vanish, and the lowest order result is essentially exact. In realistic 3DDM where $m=0$ and $\mu=0$, a magnetic field suppresses the only two other conventional orders $\Delta^s$ and $\Delta_{50}$, and leaves behind $\Delta_{5j}$ which is spin-triplet, odd-chirality, odd-parity, odd-frequency and transforms like a pseudo-vector under the Lorentz transformation. This means that $\Delta_{5j}$ can be singled-out in NMR experiment. 

Therefore the 3DDMs provide a very interesting playground for unconventional induced superconductivity in terms of pseudo-scalar or pseudo-vector, odd-frequency, and tensorial character. Even the familiar vector superconducting order parameter in 3DDM can be further classified into time-like and space-like vector superconducting orders. Understanding the interplay between various such orders and their experimental consequences~\cite{Nagai,Fuseya} requires further investigation. It maybe interesting to compare the Meissner response of the various forms of superconducting order considered here~\cite{Fominov}.
Odd-frequency triplet pairing also appears in  double quantum dots  contacted by an even-frequency s-wave superconductor  in the presence of inhomogeneous magnetic fields\cite{Bjorn}. They have  been suggested as a tool to detect unconventional pairing. The left and right dot in such a setting wouldcorrespond to left and right chirality of the present formulation.
The orders involving $\gamma^5$, break such a left-right symmetry. 

{
Recently in proximitized 3DDM, signals of $4\pi$ periodic Andreev bound states has been reported. 
Since among all the above 16 possible superconducting forms of Dirac materials, only and only $\Delta_5$ gives
rise to a non-trivial topology~\cite{Salehi2016}, this experiment is strong indication of the existence of 
pseudo-scalar superconductivity~\cite{Brinkman}. So a further prediction of our tunneling theory for the
above system is that the tensorial component of the induced superconductivity binds Majorana fermions
to a p-n junction. 
}

\section{Acknowledgements}
SAJ was supported by the Alexander von Humboldt fellowship for experienced researchers. 
ZF was supported by the research vice chancellor of Sharif University of Technology.
We thank J. K\"onig and S. Weiss for useful discussions on odd-frequency pairing.
ZF thanks M. M. Sheikh-Jabbari for supporting her visit at IPM, Tehran.


\begin{thebibliography}{99}
 The Quantum Theory of the Electron,  [2]  
\bibitem{Dirac1928}P. A. M. Dirac, Proc. R. Soc. A {\bf 117},  610 (1928).
\bibitem{Dirac_Book} P. A. M. Dirac, {\em Principles of Quantum Mechanics}, 4th ed. (Oxford University Press, Oxford, 1958).
\bibitem{ZeeBook}  A. Zee, {\em Quantum Field Theory in a Nutshell} (Princeton University Press, New Jersey, 2010). 
\bibitem{Wehling} T. O. Wehling, A. M. Black-Schaffer and A. V. Balatsky, \href{http://www.tandfonline.com/doi/abs/10.1080/00018732.2014.927109}{Adv. Phys. {\bf 63},  1 (2014). }
\bibitem{NetoRMP} A. H. Castro Neto, F. Guinea, N. M. R. Peres, K. S. Novoselov, A. K. Geim, 
\href{http://journals.aps.org/rmp/abstract/10.1103/RevModPhys.81.109}{Rev. Mod. Phys. {\bf 81},  109 (2009).}
\bibitem{ShenBook}  S. Q. Shen, {\em Topological Insulators: Dirac Equation in Condensed Matters} (Springer, 2013).
\bibitem{Tajima} N. Tajima, S. Sugawara, M. Tamura, Y. Nishio, K. Kajita, \href{http://journals.jps.jp/doi/abs/10.1143/JPSJ.75.051010}{J. Phys. Soc. Jpn. {\bf 75}, 051010  (2006).}
\bibitem{Suzumura} S. Katayama, A. Kobayashi, Y. Suzumura, \href{http://journals.jps.jp/doi/abs/10.1143/JPSJ.75.054705?journalCode=jpsj&quickLinkVolume=75&quickLinkPage=054705&selectedTab=citation&volume=75}{J. Phys. Soc. Jpn. {\bf 75},  054705 (2006).}
\bibitem{Goerbig} M. O. Goerbig, J. -N. Fuchs, G. Montambaux, F. Pi\'echon, \href{https://journals.aps.org/prb/abstract/10.1103/PhysRevB.78.045415}{Phys. Rev. B {\bf 78}, 045415 (2008)}.
\bibitem{Armitage} N. P. Armitage, E. J. Mele and A. Vishwanath, \href{https://arxiv.org/abs/1705.01111}{e-print arXiv:1705.01111 (2017).}
\bibitem{Kane_BiSO4} J. A. Steinberg, S. M. Young, S. Zaheer, C. L. Kane, E. J. Mele, and A. M. Rappe, \href{http://journals.aps.org/prl/abstract/10.1103/PhysRevLett.112.036403}{ Phys. Rev. Lett. {\bf 112}, 036403 (2014).  }
\bibitem{FuseyaReview} For a review see: Y. Fuseya, M. Ogata, H. Fukuyama, \href{http://journals.jps.jp/doi/abs/10.7566/JPSJ.84.012001?journalCode=jpsj&quickLinkVolume=84&quickLinkPage=012001&selectedTab=citation&volume=84}{J. Phys. Soc. Jpn. {\bf 84}, 012001  (2015). }
\bibitem{Kane_BiO2}  S. M. Young, S. Zaheer, J. C. Y. Teo, C. L. Kane, E. J. Mele and A. M. Rappe, \href{http://journals.aps.org/prl/abstract/10.1103/PhysRevLett.108.140405}{Phys. Rev. Lett.  {\bf 108}, 140405 (2012). }
\bibitem{WeylXu} S. Xu, I. Belopolski, N. Alidoust, M. Neupane, G. Bian, C. Zhang, R. Sankar, G. Chang, Z. Yuan, C. Lee, S. Huang, H. Zheng, J. Ma, D. S. Sanchez, B. Wang, A. Bansil, F. Chou, P. P. Shibayev, H. Lin, S. Jia and M. Z. Hasan, \href{http://science.sciencemag.org/content/349/6248/613}{Science {\bf 349}, 613  (2015). }
\bibitem{WeylDai} B. Q. Lv, H. M. Weng, B. B. Fu, X. P. Wang, H. Miao, J. Ma, P. Richard, X. C. Huang, L. X. Zhao, G. F. Chen, Z. Fang, X. Dai, T. Qian, and H. Ding, \href{http://journals.aps.org/prx/abstract/10.1103/PhysRevX.5.031013}{Phys. Rev. X  {\bf 5}, 031013 (2015).} 
\bibitem{Fang_Dai} C. Fang, M. J. Gilbert, X. Dai, and B. A. Bernevig, \href{http://journals.aps.org/prl/abstract/10.1103/PhysRevLett.108.266802}{Phys. Rev. Lett. {\bf 108}  (2012) 266802. }  
{\bibitem{Narlikar} A. V.  Narlikar (editor), {\em  Frontiers in Superconducting Materials}, (Springer-Verlag, Berlin, 2005).
\bibitem{Bennemann}  K. H. Bennemann and J. B. Ketterson (editors), {\em Superconductivity} (Springer, Berlin, 2008). }
\bibitem{Salehi} M. Salehi, S. A. Jafari, \href{http://www.sciencedirect.com/science/article/pii/S0003491615001402}
{ Ann. Phys.  {\bf  359}, 64 (2015). }
\bibitem{Fu2008PRL}
L. Fu and C. L. Kane, \href{http://journals.aps.org/prl/abstract/10.1103/PhysRevLett.100.096407}{ Phys. Rev. Lett. {\bf 100}, 096407 (2008). }
\bibitem{Salehi2016} M. Salehi, S. A. Jafari, \href{http://www.nature.com/articles/s41598-017-07298-2?WT.feed_name=subjects_materials-science}
{Scientific Reports, {\bf 7}, 8221 (2017)}
\bibitem{BohmJungYang} B. -J. Yang, N. Nagaosa, \href{http://www.nature.com/articles/ncomms5898} 
{Nat. Commun. {\bf 5}, 4898 (2014). } 
\bibitem{Morandi} G. Morandi, P. Sodano, A. Tagliacozzo, V. Tognetti, {\em Field Theories for Low-Dimensional Condensed
Matter Systems}, Springer, 2000. See chapter 6.
\bibitem{Wolff} P. A. Wolff, \href{http://www.sciencedirect.com/science/article/pii/0022369764901283}{J. Phys. Chem. Solids {\bf 25}, 1057 (1964).  }
\bibitem{gammas} Note the convention chosen in Ref. [1] is 
$\vec \gamma=\tau_2\otimes\vec i\sigma$ which differs from our choice of $\gamma$ matrrices 
by merely a rotation around $z$ axis. 
Although the form of various matrices such as $M$ used in charge conjugation may depend on the representation, 
the physics when expressed in covariant form will not depend on the representation.
\bibitem{note1} Obviously for spinful situation it is 
followed by a $i\sigma_y$ matrix multiplication for the spin components. 
\bibitem{Beenakker} C. W. J. Beenakker, \href{http://journals.aps.org/prl/abstract/10.1103/PhysRevLett.97.067007}{Phys. Rev. Lett. {\bf 97}, 067007 (2006).}
\bibitem{Belzig} W. Belzig, F. K. Wilhelm, C. Bruder, G. Schon, A. D. Zaikin, \href{http://www.sciencedirect.com/science/article/pii/S0749603699907103}{Superlatt. Microstruct. {\bf 25}, 1251 (1999).  } 
\bibitem{Goswami} P. Goswami, B. Roy, \href{https://arxiv.org/abs/1211.4023}{e-print arXiv:1211.4023 (2012).}
\bibitem{PeskinBook}  M. E. Peskin and D. V. Schroeder, {\em An Introduction to Quantum Field Theory} (Addison-Wesley Publishing Company, Massachusetts, 1995). See chapter 3.
\bibitem{blackShaffer} A. M. Black-Schaffer and A. V. Balatsky, \href{https://journals.aps.org/prb/abstract/10.1103/PhysRevB.88.104514}{Phys. Rev. B {\bf 88}, 104514 (2013). }
\bibitem{TanakaJPSJ} Y. Tanaka, M. Sato, N. Nagaosa, \href{http://journals.jps.jp/doi/10.1143/JPSJ.81.011013}{J. Phys. Soc. Jpn. {\bf 81}, 011013 (2012)}
{\bibitem{Brinkman} Ch. Li, J. C. de Boer, B. de Ronde, S. V. Ramankutty, E. v. Heumen, Y. Huang, A. Visser, A. A. Golubov, M. S. Golden and A. Brinkman, \href{https://arxiv.org/abs/1707.03154}{
e-print arXiv:1707.03154 (2017).}
\bibitem{Goerbig2009} M. O. Goerbig, J. -N. Fuchs, G. Montambaux and F. Pi\'echon, \href{http://iopscience.iop.org/article/10.1209/0295-5075/85/57005}{
Eur. Phys. Lett. {\bf 85}, 57005  (2009).}
\bibitem{machida2014} Y. Nagai, H. Nakamura and M. Machida, \href{http://dx.doi.org/10.7566/JPSJ.83.064703}
{J. Phys. Soc. Jpn. {\bf83},064703 (2014)}
}
{\bibitem{Beennaker}C. W. J. Beenakker, \href{http://www.annualreviews.org/doi/full/10.1146/annurev-conmatphys-030212-184337}{Ann. Rev. Cond. Matt. {\bf 4}, 113 (2013).}}
\bibitem{Moore} G. Y. Cho, J. E. Moore, \href{http://www.sciencedirect.com/science/article/pii/S0003491610002253}{Ann. Phys. {\bf 326}, 1515 (2011).}
\bibitem{Nagai} Y. Nagai, Y. Ota, and M. Machida, \href{http://journals.aps.org/prb/abstract/10.1103/PhysRevB.92.180502}{Phys. Rev. B {\bf 92}, 180502(R) (2015). }
\bibitem{Fuseya} Y. Fuseya, M. Ogata, H. Fukuyama, \href{http://journals.aps.org/prl/abstract/10.1103/PhysRevLett.102.066601}{Phys. Rev. Lett.  {\bf 102},  066601 (2009).  }
\bibitem{Fominov} Ya. V. Fominov, Y. Tanaka, Y. Asano, M. Eschrig, 
\href{http://journals.aps.org/prb/abstract/10.1103/PhysRevB.91.144514}{Phys. Rev. B {\bf 91}, 144514 (2015). }
\bibitem{Bjorn} B. Sothmann, S. Weiss, M. Governale,  J. K\"onig,
\href{https://journals.aps.org/prb/abstract/10.1103/PhysRevB.90.220501}{Phys. Rev. B {\bf 90}, 22051(R) (2014). }

\end{thebibliography}
\end{document}